# Nanoscale Spatial Tuning of Superconductivity in Cuprate Thin Films via Direct Laser Writing

Irene Biancardi[1], Valerio Levati[1], Jordi Alcalà[2,3], Thomas Günkel[2,3],
Nicolas Lejeune[4], Riccardo Girelli[1], Alejandro V. Silhanek[4], Valeria Russo[5], Narcís Mestres[2],
Daniela Petti[1*], Anna Palau[2*], Edoardo Albisetti[1*]

[1] Dipartimento di Fisica, Politecnico di Milano, Piazza Leonardo da Vinci 32, Milano, Italy
[2] Institut de Ciència de Materials de Barcelona, ICMAB-CSIC, Campus UAB, 08193 Bellaterra, Barcelona, Spain
[3] Departament d'Enginyeria Electrònica, Universitat Autònoma de Barcelona, 08193 Bellaterra, Barcelona, Spain
[4] Experimental Physics of Nanostructured Materials, Q-MAT, Department of Physics, Université de Liège, Sart Tilman B-4000, Liège, Belgium
[5] Dipartimento di Energia, Laboratorio Materiali Micro e Nanostrutturati, Politecnico di Milano, via Lambruschini 4, Milano, Italy

* Corresponding author.
E-mail: daniela.petti@polimi.it (DP), palau@icmab.es (AP), edoardo.albisetti@polimi.it (EA)

**Cuprate high-temperature superconductors, such as Yttrium Barium Copper Oxide (YBCO), are extremely promising for emerging technologies such as low-power computing, data storage, quantum sensors and superconducting electronics. However, the realization of high-performance functional nanostructures presents formidable challenges due to the difficulty of applying conventional nanofabrication methods to such sensitive materials, making the search for alternative methods a key enabling factor. Since YBCO's superconducting and normal-state properties are highly dependent on oxygen stoichiometry, precise nanoscale control of the oxygen content represents a highly appealing approach for creating advanced nanoengineered devices.**
**In this work, we demonstrate the precise fabrication of sub-micrometer, grayscale patterns over large areas in epitaxial YBCO thin films, achieving finely tuned optical and superconducting transport properties by locally controlling the stoichiometry through maskless direct laser writing under ambient conditions. Cryogenic magneto-optical imaging and transport measurements in irradiated devices directly demonstrate the spatial tuning of the critical temperature and carrier density with the patterning conditions. Correlated Raman microscopy and reflectometry indicate a laser-power dependent oxygen depletion in the irradiated regions.**
**The proposed laser-controlled stoichiometry approach provides a direct and scalable method to navigate the phase diagram of high-$T_C$ superconducting oxides, offering new possibilities for integrating functional nanostructures into superconducting devices.**

## INTRODUCTION

High-temperature superconductors, particularly cuprates such as Yttrium Barium Copper Oxide, $YBa_2Cu_3O_{7-\delta}$ (YBCO), play a crucial role in the development of next-generation quantum, energy and electronic technologies. With a critical temperature up to 92 K when optimally oxygenated, YBCO enables superconducting operation using liquid nitrogen, offering practical advantages over low-temperature superconductors. Its rich phase diagram comprises multiple electronic phases, including pseudogap, strange-metal, and charge-density-wave phases, some of which remain not fully understood.[1] This makes YBCO an attractive material for both fundamental research and emerging superconducting technologies.

Thanks to its high critical temperature and high critical current densities, YBCO has been employed in a variety of device applications.[2–4] For instance, YBCO-based superconducting resonators offer low-loss operation and temperature-dependent tunability, making them promising for applications in high-frequency metamaterials and sensing technologies.[5] YBCO has also been explored as a candidate for superconducting nanowire single-photon detectors (SNSPDs), offering the prospect of higher operating temperatures compared to conventional low-temperature superconductors; however, the realization of efficient SNSPDs remains limited by the difficulty of fabricating ultrathin, high-quality YBCO films.[6–8] YBCO-based Josephson junctions (JJs) have enabled the development of superconducting quantum interference devices (SQUIDs)[9–11] and exploratory demonstrations of a single-electron transistor.[12] Moreover, there has been growing interest in exploiting the complex dynamics of YBCO-based systems for energy-efficient neuromorphic computing, leveraging resistive switching and phase tunability.[13,14] While most superconducting qubits today are based on low-temperature superconductors, exploratory work is ongoing to assess the feasibility of HTS qubits that could operate at higher temperatures.[15–17] However, the widespread realization of these advanced YBCO-based devices remains significantly hindered by persistent fabrication challenges. Current nanostructuring techniques, such as Focused Ion Beam (FIB) and Ion Beam Etching (IBE), often induce damage that compromises superconducting properties, particularly in ultrathin films.[18] Maskless approaches like oxygen-ion irradiation and He-FIB have been demonstrated, offering nanometric resolution, but remain constrained by low throughput and limited scalability.[19–21] Although alternative methods are being explored, they generally involve trade-offs among resolution, scalability, and material compatibility, underscoring a critical need for advanced, non-destructive patterning strategies.

As in many complex oxides, the electronic properties of YBCO are highly sensitive to oxygen stoichiometry, which governs the transition between its various electronic phases, from antiferromagnetic insulating to superconducting. Precise control over the local oxygen content is therefore essential for the reliable integration of YBCO into functional devices. Strategies such as electromigration[22,23], electrothermal stress[24], and electric field-induced oxygen doping[25] have already demonstrated the ability to effectively tune YBCO's properties through oxygen doping. Among these, electromigration offers spatial selectivity, but it is limited to geometrical constrictions and achieves a resolution of ~1 μm. Alternative techniques that overcome these limitations and enable spatially selective tuning without geometric constraints and at submicron-scale resolution are not yet available.

Here, we present a phase nanoengineering approach: a spatially selective, non-destructive technique which allows to precisely tune the conductive and superconducting properties of YBCO over large areas while preserving its crystalline integrity.[26] Our direct-write method exploits a focused laser beam,[27–29] scanned across the material surface to induce controlled oxygen stoichiometry variations. This maskless approach offers significant advantages: it entirely circumvents the use of aggressive chemicals, complex lithographic steps, and charged particle beams, thereby mitigating the structural damage and defect formation. Furthermore, by tuning laser parameters, we demonstrate grayscale modulation of the material properties with sub-micrometer resolution, allowing for continuous and highly localized adjustment of the oxygen content and, consequently, the electronic behavior of YBCO. This capability opens a pathway to explore different regions of the YBCO phase diagram, to probe electronic phases that remain poorly understood, and to engineer enhanced functionalities in High-$T_C$ superconducting devices.

## RESULTS

### Nanopatterning of YBCO thin films

In this work, two commercial maskless laser patterning systems equipped with continuous-wave 405 nm diode lasers were employed to sequentially expose the sample surface point by point, providing a precise control of the laser power. At this wavelength, the optical penetration depth in YBCO is 55 nm (see Supporting Note 1 and **Figure S1**). A schematic of the patterning approach is shown in **Figure 1a**. The first system, characterized by a spot size of 1.2 μm and offering full control over the patterning parameters, such as the pulse duration per pixel, as well as rapid alignment with pre-existing structures, was used for systematic studies on deoxygenated YBCO films and for irradiating superconducting Hall bars. The second system, featuring diffraction-limited resolution and high-speed scanning capabilities, was utilized to investigate the ultimate throughput and spatial resolution of the laser patterning process (see Methods section).
To this end, the layout shown in **Figure 1b** was patterned on a 30-nm-thick YBCO film using a laser power of 130 mW. The pattern consists of progressively thinner lines, starting from 1 μm and decreasing down to a nominal width of 50 nm, corresponding to a single-pixel line. The Scanning Electron (SEM) image in **Figure 1c** reveals a clear and sharp contrast between the laser-written and pristine regions, with the patterned areas appearing brighter. This behavior is consistent with the effect of underdoping discussed in Section 3 and with the SEM contrast reported in the literature for other doped materials.[30] The profile extracted across the thinnest patterned lines shows a full width at half maximum (FWHM) of approximately 200 nm, which is also comparable to the minimum separation between two patterned areas, ensuring a pristine gap between them. The corresponding Electrostatic Force Microscopy (EFM) amplitude map in **Figure 1d** displays a higher surface potential in the patterned regions compared to the pristine film, indicating a locally reduced work function. The observed potential contrast thus can be linked to laser-induced oxygen depletion, resulting in a local decrease in carrier density.[31] Atomic Force Microscopy (AFM) topography maps acquired simultaneously and shown in **Figure S2** confirm that no detectable surface damage is introduced by the patterning at this level.
Then, to further test the pattern flexibility, a meander-shaped geometry was patterned at 120 mW covering a 60×180 μm$^2$ area, obtained by composing three square writing fields. The corresponding SEM image is shown in **Figure 1e**. A magnified view with the corresponding line profile is presented in **Figure 1f**, revealing that the laser-modified, less conductive regions are approximately 400 nm wide, while the unpatterned gaps are as narrow as 200 nm, demonstrating the potential of the method for fabricating superconducting nanowires. In addition, the EFM map of the same geometry in **Figure 1g** confirms the capability of the technique to realize extended structures with accurate stitching and sub-micrometer resolution.
Finally, to demonstrate the grayscale patterning and large-scale nanopatterning capability, the logo of Politecnico di Milano was written over an area as large as 500×700 μm$^2$ using four laser power levels (50, 75, 100, and 125 mW). The total patterning time for this layout was 58 s. The target layout is presented in **Figure 1h**. The corresponding SEM image in **Figure 1i** and the zoom in **Figure 1j** show distinct contrasts corresponding to different laser powers, inducing different conduction properties in the patterned material, owing to a precise multi-level spatial tuning of the stoichiometry. Last, an EFM map of the region marked in Figure 1j was acquired and is shown in **Figure 1k**: the region includes three power levels, low inside the hand, high at the border, and medium outside, corresponding to different doping levels and surface potentials. These results confirm that laser patterning enables the control of the room-temperature electronic properties with sub-micrometer spatial resolution. In the following, we study the superconducting properties of the patterned regions.

To obtain a direct spatial map of the variation in superconducting and magnetic properties after patterning, cryogenic magneto-optical imaging measurements were performed, as shown in **Figure 2**. The sample was placed beneath a bismuth-substituted yttrium iron garnet (Bi:YIG) ferrimagnetic film, which converts local magnetic fields into optical contrast via the Faraday effect. Perfectly diamagnetic superconducting regions expel the field, leaving the light polarization unchanged, whereas non-superconducting areas allow field penetration, rotating the polarization and producing a detectable contrast.

In **Figure 2a,** we show the local magnetic susceptibility at 3.3 K estimated from the cryogenic magneto-optical imaging of an array of 60×60 $\mu m^2$ square structures, fabricated via conventional optical lithography in 20 nm YBCO films. Each square structure was irradiated with a different laser power, indicated in mW below the corresponding pattern. The susceptibility maps were obtained by subtracting two images acquired at slightly different magnetic fields, thereby isolating the response of the superconducting structures to minor field variations (see Methods Section). Squares irradiated with a laser power above 89 mW are not considered in this analysis since they do not exhibit any detectable magnetic response, indicating that superconductivity is already fully suppressed at the base temperature of 3.3 K. The remaining squares, irradiated with powers below this threshold, display clear diamagnetic screening at low temperature and are sharply resolved against the background. As the temperature increases (**Figure 2b,c**, **Movie S1**), the squares irradiated with higher power within the considered range gradually lose their ability to screen the magnetic flux, indicating a loss of superconductivity. In contrast, squares irradiated with lower powers retain their superconducting properties up to nearly the critical temperature of the unpatterned YBCO (~80 K for 20-nm-thick films). **Figure 2d** shows the average susceptibility for different laser powers extracted from the magneto-optical images, plotted as a function of temperature. Squares subjected to stronger irradiation lose superconductivity at lower temperatures compared to those irradiated with lower powers. In **Figure 2e,** the critical temperature $T_c$ of each square is plotted as a function of the irradiation power. The criterion implemented to determine $T_c$ corresponds to a reduction of the susceptibility below a threshold of 25% of the low-temperature value. For reference, critical temperatures obtained from transport measurements on 15 nm-thick YBCO films are included and follow a consistent trend, validating the optical method. Further illustration of the spatial control of superconductivity is shown in **Figure 2f**, where a 260 × 100 $\mu m^2$ rectangular sample was irradiated using three different laser doses to form a meander-shaped structure. **Figures 2g-j** and **Movie S2** show the zero-field-cooled magneto-optical images of the meander under progressively increasing magnetic fields.

Magnetic flux first penetrates the regions irradiated with the maximum laser power, while the weakly irradiated and unpatterned sections remain in the superconducting state. As the applied field is further increased, flux progressively penetrates larger portions of the patterned regions. This sequential penetration reflects the spatial variations in the magnetic flux penetration field $H_P$, associated with the locally reduced critical temperature.[32] Eventually, at the highest applied field, superconductivity is preserved only in the unpatterned regions, which form well-defined superconducting paths along the meandering geometry.

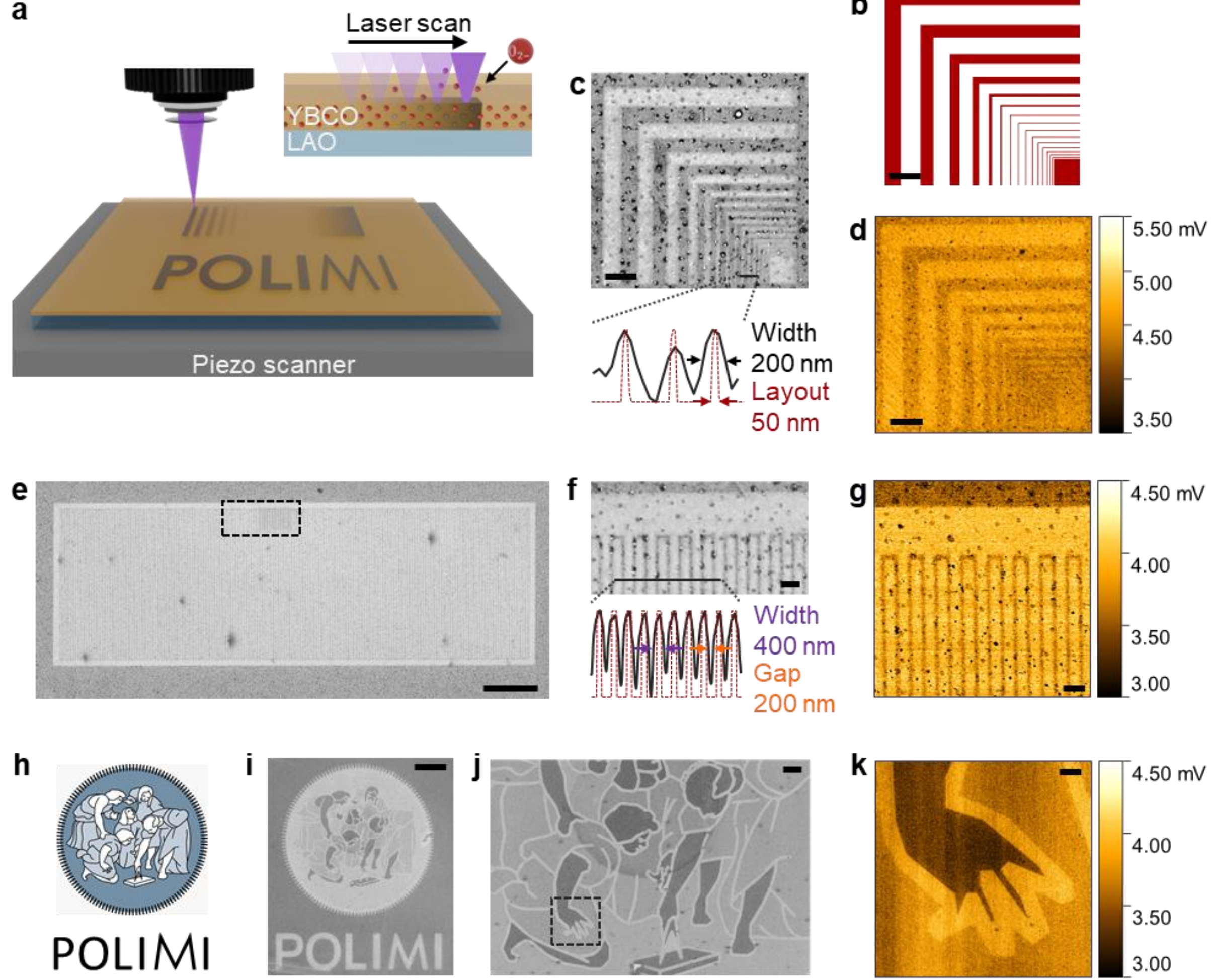

**Figure 1**. **Laser driven local oxygen doping in YBCO thin films**.
(**a**) Schematic illustration of the patterning process. A focused CW 405 nm laser is raster scanned on a YBCO film in ambient conditions, producing a local deoxygenation of the material. (**b-d**) Test pattern (12×12 $\mu m^2$) with progressively thinner lines along both in-plane directions. Scale bars: 2 µm. (**b**) Target layout. (**c**) SEM image of the structure corresponding to the layout in panel (**b**) and extracted line profile, showing a minimum feature size of ~200 nm. (**d**) Electrostatic Force Microscopy (EFM) map of the layout in panel (**b**), revealing a higher surface potential in the patterned area compared to the pristine film. (**e-g**) 60×180 $\mu m^2$ meandering structure spanning three stitched writing fields. (**e**) SEM image of the whole pattern. Scale bar: 10 µm. (**f**) SEM image of the magnified view with corresponding line profile, showing a minimum gap of ~200 nm between adjacent features, and (**g**) EFM image. Scale bars: 1 µm. (**h-k**) Large scale nanopattern of the Politecnico di Milano logo (500×700 $\mu m^2$) featuring four grayscale levels achieved by varying the laser power: (**h**) layout design, (**i**) SEM image (scale bar: 100 µm), (**j**) magnified view (scale bar: 20 µm), and (**k**) EFM image of the hand detail (scale bar: 5 µm).

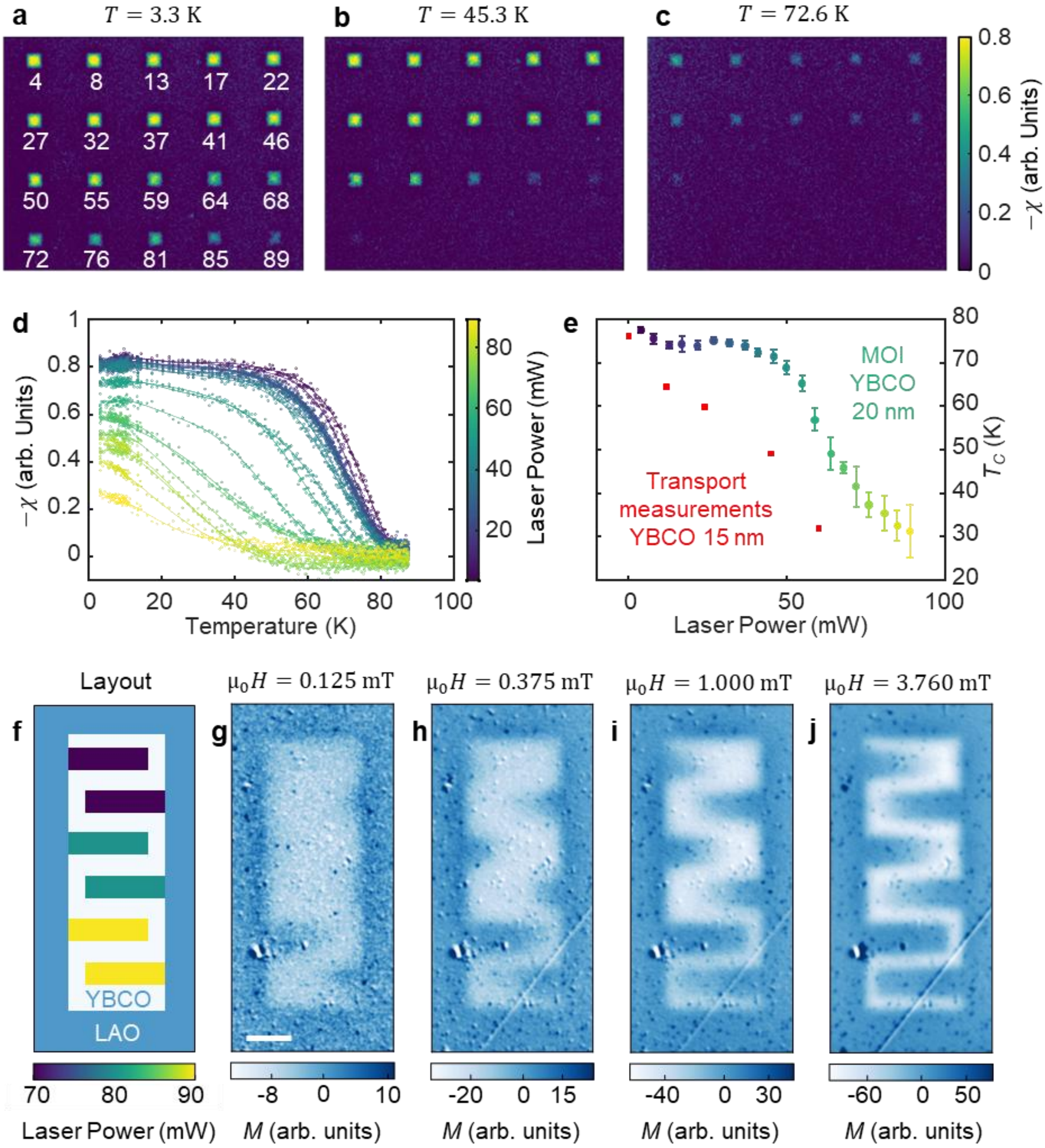

**Figure 2. Magneto Optical Imaging.**
(**a–c**) Magneto-optical images of the patterned sample where the color code is related to the local magnetic susceptibility. The sample was prepatterned with 60×60 μm² YBCO squares that were subsequently laser irradiated with power (in mW) indicated in panel (**a**). The susceptibility is obtained by subtracting two images acquired at slightly different magnetic field values, thus isolating the sample's response to the field variation (see Methods for details). At low temperature (**a**), all squares exhibit diamagnetic behavior and are clearly visible. Upon increasing temperature (**b,c**), the squares irradiated with higher dose progressively lose their screening property. (**d**) Average susceptibility extracted from the magneto-optical images as a function of laser power and temperature. Strongly irradiated squares become non-superconducting at lower temperatures compared to weakly irradiated ones, whose critical temperature remains close to that of unpatterned YBCO ($T_c \approx 80$ K). (**e**) Critical temperature $T_c$ of each square as a function of the laser power. For comparison, $T_c$ estimated from transport measurements on 15 nm YBCO films (discussed in Figure 4) are shown in red. (**f**) Example of a 260×100 μm² rectangle patterned in a meander shape using three different laser powers. (**g-j**) Zero-field-cooled magneto-optical images of the meander under increasing magnetic field. Flux penetration starts in the highly irradiated regions, while the unpatterned ones remain superconducting, compatible with a reduced critical field in the patterned areas. Scale bar: 50μm.

**Optical and structural characterization**

To further investigate the optical and structural properties of pristine and modified YBCO we performed reflectometry and Raman spectroscopy analyses on a 100-nm-thick sample.

**Figure 3a** shows an optical image of a series of 50×50 $\mu m^2$ patterned squares, irradiated with increasing laser power and a fixed laser pulse duration equal to 25 µs. The patterned regions appear brighter, with contrast that becomes more pronounced at higher laser power. This enhanced visibility is attributed to laser-induced modifications of the material's optical properties caused by local oxygen deficiency, consistent with previous studies. [33] Reflectance spectra in **Figure 3b** and **Figure S3** support this interpretation: a peak emerges at 4.1 eV exclusively in the laser-modified regions, monotonically increasing and shifting with laser power. This feature is commonly associated with the formation of $Cu^{1+}$ in oxygen-deficient O(1)–Cu(1)–O(1) chains and serves as a fingerprint of reduced oxygen content in YBCO.[34–36] Micro-Raman spectra acquired from the same set of patterns are presented in **Figure 3c**. The primary spectral features remain preserved across all squares, indicating that the crystallinity of YBCO is not significantly altered within the investigated range of irradiation powers. The most notable difference is the reduction and shift toward lower wavenumbers of the 500 $cm^{-1}$ mode. This mode corresponds to the vibrational motion of the apical oxygen O(4), which is bonded to Cu(1) and is therefore sensitive to deoxygenation.[37–39] To precisely determine the shift of the O(4) peak while accounting for all contributing features, the spectral region around the 500 $cm^{-1}$ mode was fitted using three Lorentzian components: the O(4) vibration, a fixed peak at 486 $cm^{-1}$ due to the LAO substrate, and a less intense and broader contribution at lower frequencies associated with O(2,3) modes.[40] An example of the fitting outcome is shown in **Figure 3d** for the pristine spectrum. The results of this analysis, shown in **Figure 3e** as a function of the power, confirmed the systematic shift of the O(4) peak from 501 $cm^{-1}$ in the pristine case to 498 $cm^{-1}$ at the maximum power studied. Moreover, from the Raman shift, it was possible to estimate the oxygen stoichiometry of each pattern (see Supporting Note 2). For the $YBa_2Cu_3O_x$ pristine film, this analysis yields an oxygen content of $x \approx 6.80$, slightly below optimal doping, whereas irradiation at the maximum laser power of 34 mW reduces the oxygen content to $x \approx 6.68$. According to the experimental data reported in Ref. [41], this oxygen concentration corresponds to a critical temperature of approximately 64 K. This value is higher than the minimum $T_c$ reported in Figure 2e, as inferred from MOI on a 20-nm-thick sample. Indeed, the larger film thickness required to ensure a sufficient Raman signal results in higher optical absorption and, consequently, in the onset of topographic surface degradation before largely deoxygenated states can be reached. Conversely, laser irradiation of thinner samples allows access to lower values of the critical temperature, and thus to a wider range of oxygen stoichiometry. Overall, these results confirm that oxygen depletion is the dominant mechanism governing the laser-induced variations in the electrical properties discussed in the previous section. This depletion is likely driven both by thermally activated migration due to localized annealing during laser writing and by the effects induced directly by UV irradiation.[42,43]

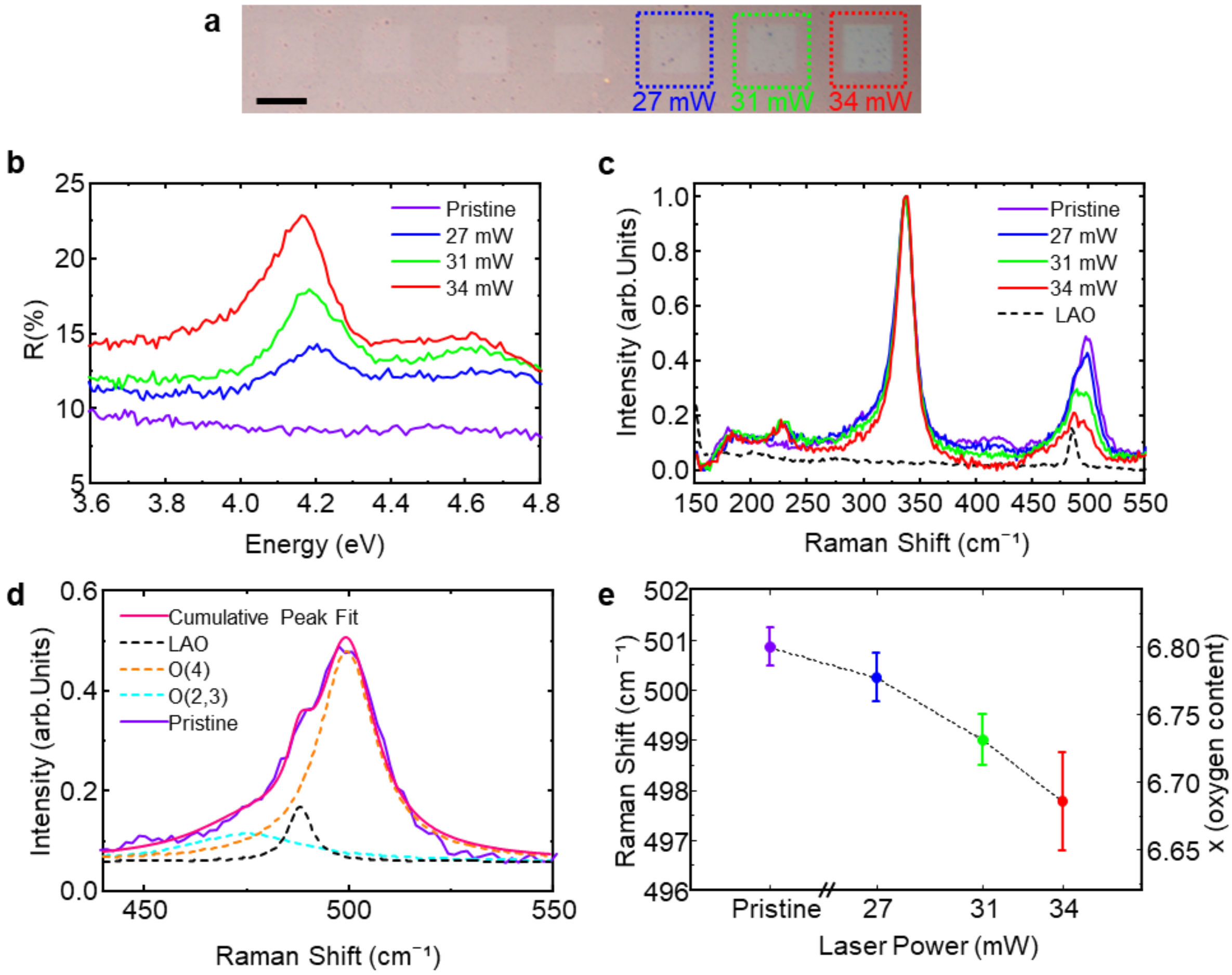

**Figure 3**. **Structural and optical characterization.**
(**a**) Optical image of 50×50 μm² squares, patterned with increasing laser power from left to right and laser pulse duration of 25 μs. Scale bar: 50 μm. (**b**) Reflectance spectra of the squares framed in panel (**a**), showing the emergence of a characteristic peak around 4.1 eV exclusively in the patterned areas, with intensity proportional to the laser power, fingerprint of deoxygenation. (**c**) Raman spectra acquired on the same set of patterns, highlighting a reduction and shift of the apical oxygen peak near 500 cm⁻¹ with increasing patterning power. (**d**) Peak fitting in the 420–550 cm⁻¹ range for the pristine spectrum. The signal is separated into three components, corresponding to the O(4) mode, a contribution from the LAO substrate, and a broader feature associated with the O(2,3) mode. (**e**) Raman shift of the O(4) mode calculated from the fitting and corresponding oxygen content as a function of the laser power used for patterning, the error value is the fitting uncertainty. A shift from 501 cm⁻¹ corresponding to the pristine to 498 cm⁻¹ for the maximum laser power is observed, linked to a decrease in oxygen stoichiometry from 6.8 to 6.68.

**Tuning of the superconducting transport properties**

To complement the previous characterization, cryogenic electronic transport measurements were performed on Hall-bar devices with irradiated channels to quantitatively assess the electrical response of the laser-written YBCO patterns. The device geometry and patterning procedure are illustrated in **Figure 4a**.

As an initial step, we characterized channels uniformly exposed to different laser powers, ranging from 12 mW to 60 mW, on a 15-nm-thick sample. A representative optical microscopy image of the channel is shown in **Figure 4b**, where the patterned region appears brighter, consistent with the results presented in the previous section. The dependence of the critical temperature on the patterning laser power was investigated by measuring resistance as a function of temperature, as shown in **Figure 4c**. A first result is that all devices exhibit the characteristic superconducting transition. Most importantly, it is shown that the superconducting critical temperature can be continuously tuned by laser irradiation, gradually decreasing as the laser power is increased. This controlled modulation of $T_c$ is consistent with the magneto-optical characterization presented in Figure 2. In parallel, a slight broadening of the transition is also observed at higher power and is appreciable in the logarithmic representation in **Figure 4d**. This behavior is commonly observed in YBCO subjected to deoxygenation by other methods, e.g. electric-field induced oxygen doping.[25] Furthermore, the room-temperature resistance monotonically increases with increasing laser power, indicating that irradiation affects both superconducting and normal-state transport. This correlation is crucial to confirm that the SEM and EFM investigation of the patterned regions at room temperature (Figure 1), can be useful and straightforward indicators of the low-temperature transport behavior. Notably, these transport properties remained unchanged when the measurements were repeated after one month, confirming the stability of the laser-induced modifications over this time scale.

To quantify the variation of the critical temperature, we extracted $T_c$ as the temperature at which the resistance falls to 0.1% of its value at $T = 300$ K. The results, reported in **Figure 4e** for both the 15-nm-thick film (corresponding to the curves in Figure 4c,d) and a 20-nm-thick film ($R$-$T$ curves not shown here), indicate a consistent monotonic reduction of the critical temperature with increasing laser power for both thicknesses. Specifically, at the maximum laser power, $T_c$ lowers from 76.2 K to 32 K for the 15-nm-thick sample and from 82 K to 37 K for the 20-nm-thick one. Furthermore, by increasing the laser power beyond these values, it is also possible to achieve a complete suppression of superconductivity, as illustrated by the resistance versus temperature measurement reported in **Figure S4**.

Next, we evaluated the influence of laser irradiation on the critical current density as a function of the temperature, in the same devices. In this analysis, $J_c$ is defined as the current density corresponding to a voltage drop across the bridge of 20 nV. As shown in **Figure 4f**, the self-field $J_c$ systematically decreases with increasing laser power. At higher patterning powers, the suppression of $J_c$ becomes much more pronounced: the self-field $J_c$ drops from 2.70 MA/cm$^2$ at 24 mW to 0.89 MA/cm$^2$ and 0.34 MA/cm$^2$ at 45 mW and 60 mW, respectively.

Subsequently, the magnetic-field dependence of $J_c$ was measured at a fixed temperature of 30 K for tracks irradiated at 24 mW and 45 mW, along with a non-patterned one. As shown in the log-log plot of **Figure 4g**, the three curves start from a different self-field $J_c$ values and remain nearly constant for magnetic fields up to about 0.1 T. This region, associated with single-vortex pinning, indicates a homogeneous deoxygenation process without introducing additional defects such as stacking faults, which would otherwise extend the $J_c(H)$ plateau.[25] At higher fields, all curves rapidly decrease to zero as they approach the irreversibility line.

Finally, Hall measurements were performed at room temperature (300 K) to estimate the carrier density $n_H$ in the irradiated regions (see **Figure S5**). The phase diagram of YBCO was reconstructed, correlating $n_H$ with the critical temperature and the pseudogap temperature, determined from the deviation of the high-temperature linear behavior of the resistance. The data for both 15-nm-thick and 20-nm-thick samples are reported in **Figure 4h**. Laser patterning clearly shifts the material towards lower doping levels, enabling precise control over progressively deoxygenated states. The observed trends align well with the expected phase diagram, demonstrating that laser patterning provides an effective approach to continuously tune the doping level and access different regions of the YBCO phase diagram.

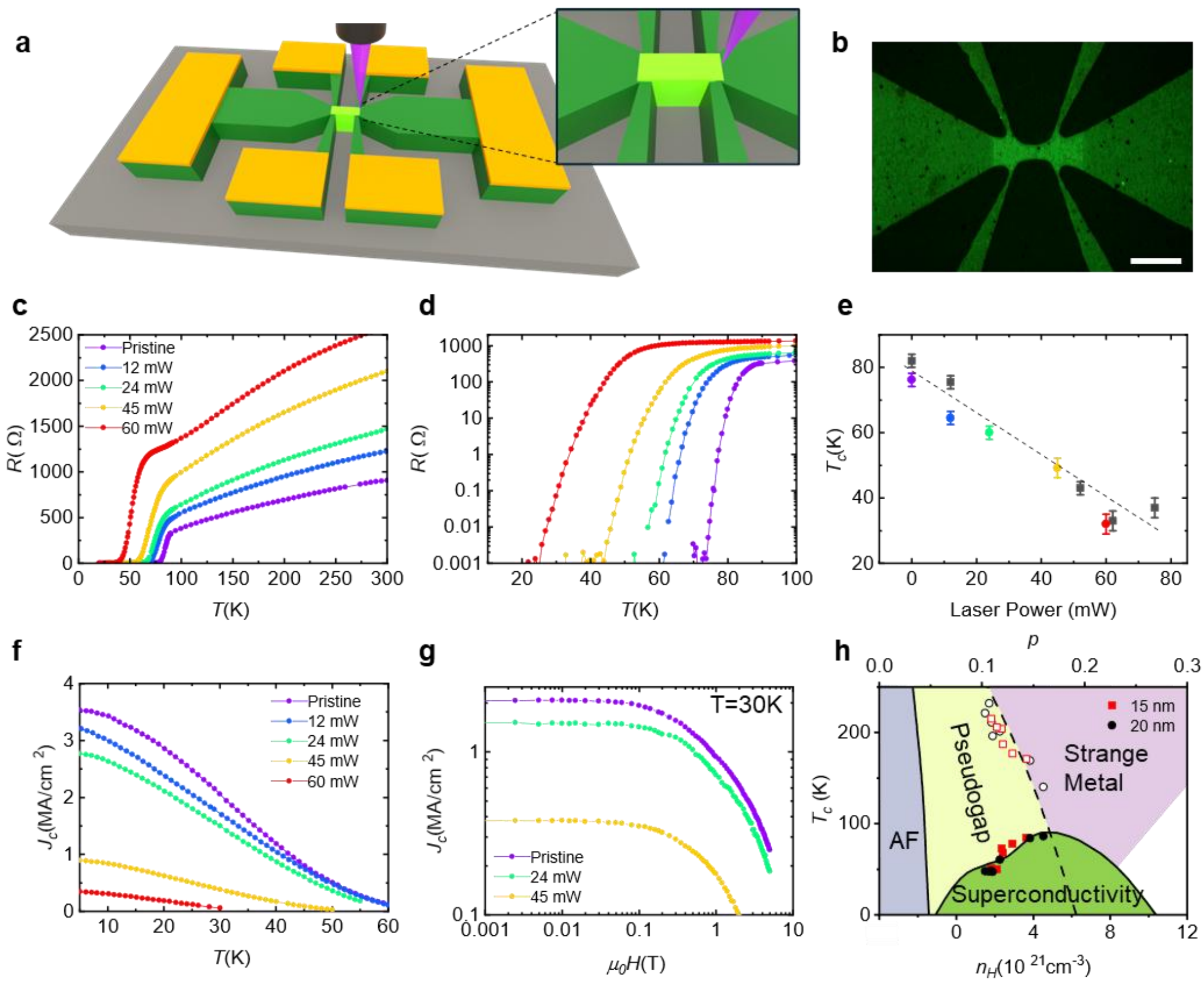

**igure 4. Tunable superconducting properties in irradiated Hall bar devices.**
(**a**) Schematic representation of the laser irradiation of the channel of a YBCO Hall bar with gold pads for electrical probing. (**b**) Optical image of the channel after laser exposure, where the uniformly patterned area appears as a brighter region. To enhance its visibility, the image was digitally filtered on the green component. Scale bar: 20 μm. (**c,d**) Resistance-temperature curves for devices fabricated on the 15-nm-thick sample: four irradiated at different laser powers and fixed pulse time of 100 μs along with pristine channel reference. All curves display the characteristic shape of phase transition into the superconducting state, whose value of critical temperature depends on the laser power used for patterning. The data are plotted in linear (**c**) and logarithmic (**d**) scales. (**e**) Critical temperature as a function of irradiation power. Squares correspond to data from the 15-nm-thick sample, while circles refer to the 20-nm-thick one (data not shown here). Error bars correspond to the temperature sweep step size. (**f**) Critical current as a function of temperature for the pristine and irradiated devices shown in panel (**c**). (**g**) Critical current as a function of the applied perpendicular magnetic field at 30 K for the devices irradiated at 24 mW and 45 mW, compared to the pristine one. (**h**) Phase diagram of YBCO showing the experimental critical temperatures and pseudogap temperatures as functions of the calculated carrier density extracted from Hall measurements on the same devices. The experimental data are in good agreement with the theoretical phase diagram of YBCO.

**Spatially dependent control of superconductivity**

After studying the modifications induced by uniform laser patterning, we explored the potential for spatially modulated irradiation within the superconducting tracks by patterning two demonstrative layouts with a 1.2 µm spot-size laser.

The first geometry, shown in the optical image and the schematic in **Figure 5a**, consists of a constriction that narrows the superconducting channel to widths ranging from 2 µm down to 0.75 µm. Resistance-temperature characteristics were measured for the series of constrictions and compared to a pristine 10-µm-wide track. The results are shown in **Figure 5b** in logarithmic scale. The resistance curve of the pristine device nearly overlaps with that of the 2-µm-wide constriction, indicating that the track was accurately patterned while fully preserving its superconducting properties. For narrower constrictions, a progressive reduction of the critical temperature is observed, reaching ~20 K for the narrowest device. In all patterned tracks, a first transition occurs at the same critical temperature as the pristine reference, consistent with the contribution of the unmodified pristine material within the tracks. **Figure 5c** shows the dependence of the self-field $J_c$ on constriction width. Consistent with the $T_c$ trend, tracks narrower than 2 µm exhibit partial deoxygenation likely due to a partial irradiation of the central region, consistently with the expected 1.2 um spot size. The narrowest constriction shows a reduction in $J_c$ by a factor of 5 at 5 K. These results confirm that sub-micrometer tracks can be patterned using this approach; in this framework, a sizeable enhancement of the spatial resolution and minimum constriction width is expected using a diffraction limited spot size and by further optimizing the patterning parameters.

The second geometry, presented in **Figure 5d**, features several transversal lines patterned across the same track, irradiated either with the same power, or with different laser powers. The resistance vs temperature measurements of **Figure 5e** show the curves, normalized to $R(100\ \mathrm{K})$, compared to those of the pristine device. In all the exposed tracks, a first transition appears at the critical temperature of the pristine sample, followed by subsequent transitions at lower temperatures. To better resolve these transitions, the temperature derivative of the resistance is plotted in **Figure 5f**, revealing different peaks associated to regions with different local oxygen content. For lines irradiated uniformly at 77 mW, the resistance curve displays two well-separated superconducting transitions. The first one, at $T_{c,1} \approx 85$ K, corresponds to the unmodified region, while the second one, at $T_{c,3} \approx 45$ K, originates from the laser-exposed region. When the lines are patterned using multiple irradiation powers (42, 53, 65 and 77 mW), the two transitions described above remain visible. In addition, a broader intermediate transition appears at $T_{c,2} \approx 65$ K, arising from the portions irradiated at the lower powers (42, 53, and 65 mW). Finally, when the track is patterned at five powers (42, 53, 65, 77, and 86 mW), the resistance curve develops a fourth transition at $T_{c,4} \approx 35$ K, relative to the area patterned at the highest power. This behavior reflects the coexistence of regions with different irradiation-induced transport properties. Under the present experimental conditions, the derivative analysis therefore allows us to resolve at least four distinct superconducting levels within a single device, associated with the pristine region and with regions exposed to different irradiation powers.

Overall, these results establish direct laser writing as a tool for spatially programming superconducting electronic functionalities, enabling selected micrometric regions within the same device to be patterned with distinct oxygen doping profiles.

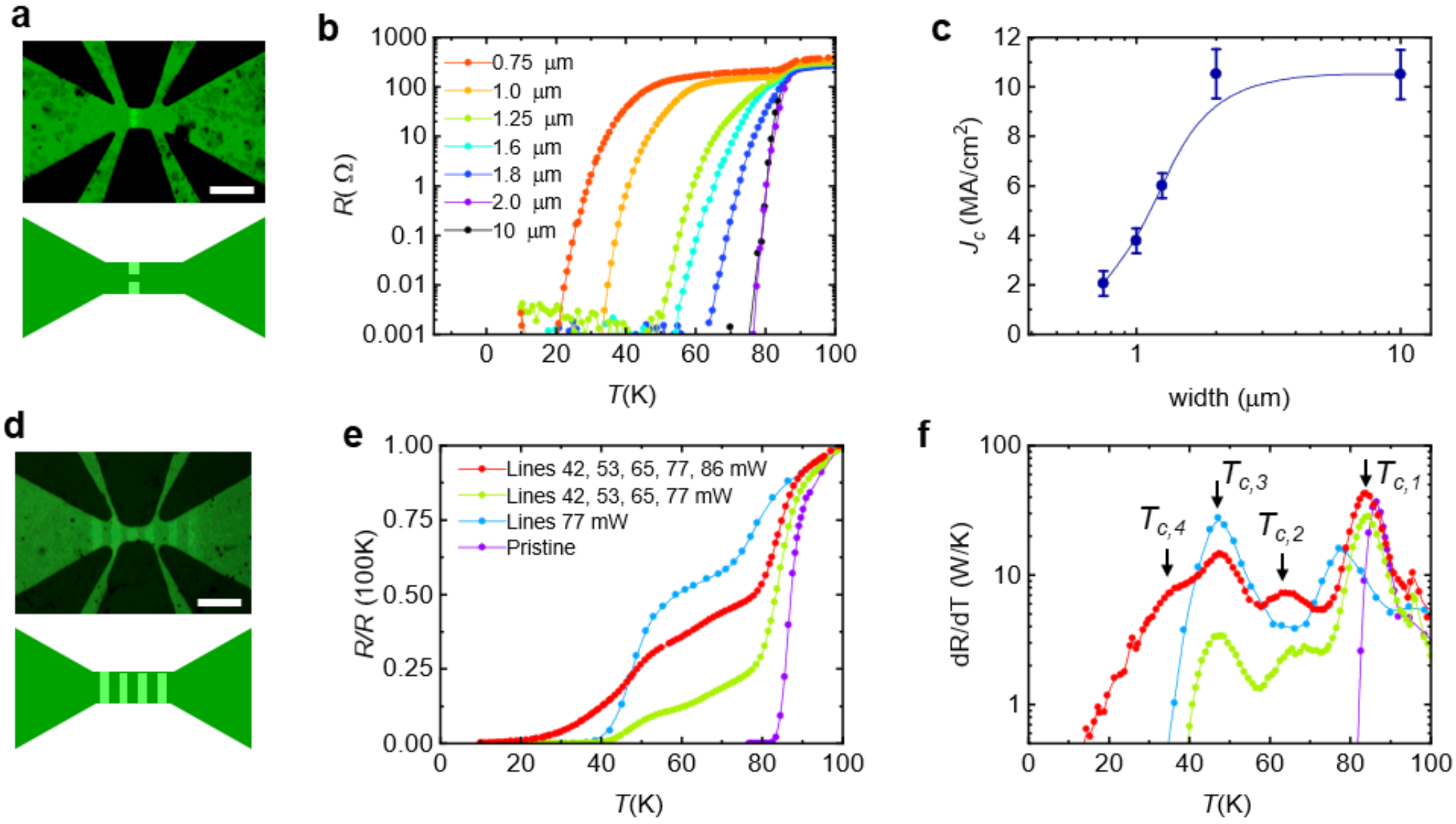

**Figure 5**. **Spatially-resolved control of the critical temperature.**
(**a**) Optical image, digitally filtered on the green component, together with a scheme of the device track patterned with a constriction geometry leaving a ~1 μm-wide conducting gap. Scale bar: 20 μm. (**b**) Resistance as a function of temperature, plotted on a logarithmic scale, for devices patterned with constrictions with different widths and fixed power (95 mW) and pulse time (25 μs) on a 20-nm-thick sample. (**c**) Self-field $J_c$ as a function of the width of the constrictions. (**d**) Optical image, digitally filtered on the green component, and corresponding schematic of a device track patterned with transversal lines 3 μm thick. Scale bar: 20 μm. (**e**) Resistance as a function of temperature, normalized to R(100 K), for devices patterned with transversal lines on a 20-nm-thick sample with 25 μs of pulse time. (**f**) Derivative of the *R(T)* curves in panel (**e**), highlighting the positions of the different critical temperatures associated with different level of local deoxygenation.

## CONCLUSION

Direct laser writing emerges as a powerful and versatile tool for the fabrication of YBCO-based superconducting devices, offering unprecedented control over local oxygen doping. This approach not only enables patterning at the sub-micrometer scale but can also be efficiently applied to large areas, allowing rapid fabrication of extended structures. At the same time, it allows fine-tuning of the carrier concentration within the patterned regions, thereby modulating normal-state and superconducting properties with remarkable precision both spatially and stoichiometrically. In this way, it enables studies of transport phenomena and material interactions across a broad portion of the YBCO phase diagram, without the need to optimize different growth conditions. Changes in oxygen content directly affect the critical temperature and critical current, in addition to local magnetic-field penetration and optical responses, making it possible to create areas with spatially varying functionalities. Importantly, this technique is compatible with structurally delicate cuprate materials and, unlike conventional nanofabrication approaches that produce only binary modifications, direct laser writing allows continuous "grey-scale" modulation of material properties under ambient conditions, offering a straightforward, single-step, route to engineer complex superconducting architectures. In perspective, this approach can be further explored to achieve reversibility, for example by annealing in an oxygen-rich atmosphere, and to directly write superconducting paths starting from a deoxygenated film at low temperature.

Overall, the versatility of the approach opens new perspectives not only for YBCO-based devices but also for other cuprates and perovskite oxides, providing a scalable platform to integrate functional nanostructures and investigate novel phenomena in next-generation oxide technologies.

## METHODS

*Direct laser Writing*: Laser patterning was carried out using two complementary systems by Heidelberg Instruments: NanoFrazor Explore and a DWL 66+, both equipped with a continuous-wave (CW) 405 nm semiconductor diode laser.
For the NanoFrazor Explore system, the laser beam, with a nominal spot size of 1.2 μm, was focused onto the sample surface using a 20× objective. Patterns were defined as arrays of 50×50 $nm^2$ pixels, and the sample was scanned in a raster fashion by a piezoelectric stage at a constant velocity determined by the pixel size and the pixel time, which ranged from 55 μs to 130 μs in this work (see Figure S6 for details on the effect of irradiation time variation). During each pixel time, the laser was turned on for a duration defined by the pulse time, equal to the pixel time minus 30 μs. As an example, for a pixel time of 55 μs and a pulse duration of 25 μs, the total patterning time required to irradiate a 50×50 $\mu m^2$ area was 138 s. This tool was employed for the fabrication of the patterns reported in Figure 2-5.
For the DWL 66+ system, the High-Resolution Mode was employed, providing a minimum feature size of 200 nm. The system automatically focused the laser beam onto the sample surface, and the scan speed was fixed at 1800 mm $s^{-1}$. A 50×50 $nm^2$ pixel grid was used, resulting in a total patterning time of 7 s for a 50×50 $\mu m^2$ square. Crucially, the 50 nm step size of the writing grid ensures a high degree of spatial overlapping. The spatial resolution achieved on the film is therefore dictated by the interplay between these writing conditions and the material response. In particular, when operating close to the modification threshold, only the central region of the Gaussian beam profile contributes significantly to the process, enabling effective linewidths down to 200 nm under optimized exposure conditions. This tool was employed for the fabrication of the patterns reported in Figure 1.

*Electrostatic Force Microscopy*: EFM maps were acquired using a Park Systems NX10 AFM operating in non-contact mode with PPP-EFM probes by Nanosensors. To ensure proper electrical contact, the sample was mounted on a conductive disk using silver paint, and 2 V were applied to it during measurements.

*Scanning Electron Microscopy*: SEM images were acquired using a Leo/Zeiss 1525 FE-SEM equipped with in-lens detector for surface specific imaging. The measurements were performed with an accelerating voltage of 3 kV and a beam current of 130 pA.

*Magneto-Optical Imaging:* The magnetic flux landscape was directly visualized using magneto-optical imaging (MOI). This technique exploits the Faraday rotation of linearly polarized light within a three μm-thick Bi-doped yttrium iron garnet indicator film that has in-plane magnetic domains and is placed in direct contact with the sample. Because polarization rotation scales with the local out-of-plane magnetic field component, an analyzer set in crossed Nicols configuration relative to the polarizer produces images whose intensity is proportional to this component. Image acquisition was performed with a CCD camera. Post-processing, conducted using Python, corrected for uneven illumination and background contributions independent of the field. A detailed description can be found in Ref [44]. Low-temperature measurements were carried out in a closed-cycle cryostat with a cooling power of approximately 100 mW at 4 K. The external magnetic field was applied using a copper coil with an experimentally determined sweep rate of about 44 mT/s. MOI thus enables the creation of spatially resolved flux maps, providing direct information on the magnetic flux distribution as well as the position and size of magnetic avalanches. MOI also allows for magnetic susceptibility imaging, as shown in Figure 1. In this case, the sample was field cooled under 1.6 mT down to the base temperature and then the temperature was increased while recording images at two applied fields, 2.4 mT and 0.8 mT. For each temperature, the image difference between these two fields was calculated, which is proportional to the magnetic susceptibility $\chi$ of the YBCO.

*Optical reflectance measurements*: The reflectance spectra were collected using a Filmetrics F54-XY-UVX system equipped with a 10× objective.

*Raman Spectroscopy*: Raman spectra were acquired using a Renishaw InVia micro-Raman spectrometer with a 457 nm Argon ion laser as the excitation source. The laser was focused onto the sample surface through a 50× objective, delivering an effective power of 3 mW. Each spectrum consisted of 251 uniformly spaced data points spanning the 160–550 $cm^{-1}$ range. An acquisition time of 10 seconds was used per each spectrum, and the final spectra were obtained by averaging 100 repetitions to enhance the signal-to-noise ratio.

*Electronic Transport Measurements*: Cryogenic magneto-transport experiments were carried out in a Quantum Design Physical Property Measurement System (PPMS). The instrument provides precise temperature control down to ≈ 2 K and magnetic fields up to 9 T, with field stability better than 0.1%. Measurements were performed in AC Transport mode, using a 0.01 mA excitation current at 33 Hz to acquire four-probe resistance and magnetotransport data on microfabricated YBCO Hall bar devices. Before each measurement session, the samples were mounted on a PPMS puck with copper tape to ensure efficient thermal anchoring. Electrical contact between the devices (two per puck) and the measurement leads was achieved by ultrasonic wire bonding (Kulicke & Soffa 4526) using aluminum wires.

**ACKNOWLEDGEMENTS**

This work was partially carried out at PoliFab, the microtechnology and nanotechnology center of the Politecnico di Milano. V.L. acknowledges Matteo Fettizio for his valuable assistance with wire bonding. E.A. acknowledges funding from the European Union's Horizon 2020 research and innovation programme under grant agreement number 948225 (project B3YOND). N.L. acknowledges support from FRS-FNRS (Research Fellowships FRIA).
A.V.S acknowledge financial support from The Research Foundation - Flanders (FWO) and the F.R.S.-FNRS through the Excellence of Science (EOS) programme (project number O.0028.22), from COST (European Cooperation in Science and Technology) through COST Action SUPERQUMAP (CA21144), and from the European Union's Horizon 2020 research and innovation programme under Grant agreement number 101007417 having benefited from the access to the facilities of ICN2 - Fundació Institut Català de Nanociència i Nanotecnologia, Barcelona (ES), within the framework of the NFFA-Europe Pilot Transnational Access Activity, proposal ID557. T.G. acknowledges support from the AGAUR Catalan Government through a Predoctoral Fellowship (2022 FISDU 00115). A.P. and N.M. acknowledge financial support from the Spanish Ministry of Science and Innovation (MCIN/AEI/10.13039/501100011033) through the "Severo Ochoa" Programme for Centres of Excellence (CEX2023-001263-S), the projects HTSUPERFUN (PID2021-124680OB-I00) and HTS-4ICT (PID2024-156025OB-I00), co-financed by ERDF - A way of making Europe, and from the Spanish Nanolito networking project (RED2022-134096-T). A.P and N.M. also acknowledge the Scientific Services at ICMAB and the PhD programme in Materials Science at the Universitat Autònoma de Barcelona (UAB).

**AUTHOR CONTRIBUTIONS**

I.B., and V.L. performed the laser nanopatterning. I.B. and V.L. performed the EFM characterization. I.B. performed the SEM measurements. A.V.S and N.L. performed the magneto-optical imaging and data analysis. I.B and V.L. performed the structural and optical characterization. I.B., V.L. and V.R. performed the Raman measurements. J.A. and T.G. fabricated the Hall bars. J.A., T.G., V.L. and A.P. performed the electronic transport measurements and data analysis. I.B., V.L., A.V.S., A.P. and E.A. wrote the manuscript aided by T.G., N.L., N.M. and D.P. with contributions from all the authors. All authors contributed to discussion and data interpretation. A.P., A.V.S., D.P. and E.A. designed and supervised the research.

**COMPETING INTERESTS**

The authors declare no competing interests.

**Supporting Information**

# Supporting Information for: Nanoscale Spatial Tuning of Superconductivity in Cuprate Thin Films via Direct Laser Writing

Irene Biancardi, Valerio Levati, Jordi Alcalà, Thomas Günkel,
Nicolas Lejeune, Riccardo Girelli, Alejandro V. Silhanek, Valeria Russo, Narcís Mestres,
Daniela Petti*, Anna Palau*, Edoardo Albisetti*

## Supporting Note 1

The laser beam profile and the corresponding intensity attenuation across the film thickness were estimated following the approach reported in the Supplementary Information of Ref [1].
A Gaussian beam profile has been assumed, with spatially varying intensity given by:

$$I(x,y,z) = \frac{2P_0}{\pi w^2(z)} e^{-2\frac{x^2+y^2}{w^2(z)}}$$

where $P_0$ is the incident laser power and $w(z)$ is the beam waist along the propagation direction.

The system has been modeled as a multilayer structure composed of air ($z < 0$), YBCO ($-d < z < 0$) and LAO ($z < -d$). The beam propagation in each medium is treated by introducing an effective wavelength $\lambda/n$, which modifies the corresponding Rayleigh range. Within this approximation, the beam waist evolution in each layer can be written as:

$$w_{VAC}(z) = w_0 \sqrt{1 + \left(\frac{z - z_0}{z_{R,VAC}}\right)^2}, \qquad z_{R,VAC} = \frac{\pi w_0^2}{\lambda};$$

$$w_{YBCO}(z) = w_0 \sqrt{1 + \left(\frac{z - z_{0,YBCO}}{z_{R,YBCO}}\right)^2}, \quad z_{R,YBCO} = \frac{n_{YBCO} \pi w_0^2}{\lambda}, \quad z_{0,YBCO} = z_0 n_{YBCO}$$

$$w_{LAO}(z) = w_0 \sqrt{1 + \left(\frac{z - z_{0,LAO}}{z_{R,LAO}}\right)^2}, \qquad z_{R,LAO} = \frac{n_{LAO} \pi w_0^2}{\lambda}, \qquad z_{0,LAO} = z_0 n_{LAO} + d \frac{n_{LAO}}{n_{YBCO}} - d$$

The refractive index of YBCO at 405 nm is $n_{YBCO} = 1.32$.[2] The resulting intensity profile along the beam propagation direction has been calculated and is reported in the **Figure S1a** in the case of $d = 100\ nm$, which corresponds to the thicker film used in this study.

Considering the Gaussian beam intensity along the optical axis ($x = 0, y = 0$) and adopting an effective Lambert-Beer attenuation model, the intensity decay inside the YBCO layer can be expressed as:

$$I_{YBCO} = \frac{2P_0(1-R)}{\pi w_{YBCO}^2(z)} e^{-\alpha_{YBCO}|z|}$$

where $R$ is the Fresnel reflection coefficient at the air/YBCO interface, accounting for the complex refractive index of YBCO:

$$R = \frac{(n_{YBCO} - 1)^2 + \kappa^2}{(n_{YBCO} + 1)^2 + \kappa^2}$$

and the absorption coefficient is given by:

$$\alpha_{YBCO} = \frac{4\pi\kappa}{\lambda} = 18 * 10^4 cm^{-1}$$

with the extinction coefficient $\kappa = 0.6$.[2] This corresponds to an optical penetration depth of:

$$\delta_{YBCO} = \frac{1}{\alpha_{YBCO}} = 55\ nm$$

Back-reflections at the YBCO/LAO interface have been neglected, since their contribution is expected to be small compared to the strong absorption within the YBCO layer at 405 nm.

The resulting intensity profile along the propagation direction has been evaluated for different incident laser powers and is shown in **Figure S1b**.

Therefore, the 15-20 nm thick films used for cryogenic transport measurements and magneto-optical imaging (MOI) are thinner than the optical penetration depth and the optical intensity can be considered approximately uniform across the film thickness. In contrast, for the 100-nm-thick films used for Raman and reflectance characterization, an intensity gradient develops across the thickness.

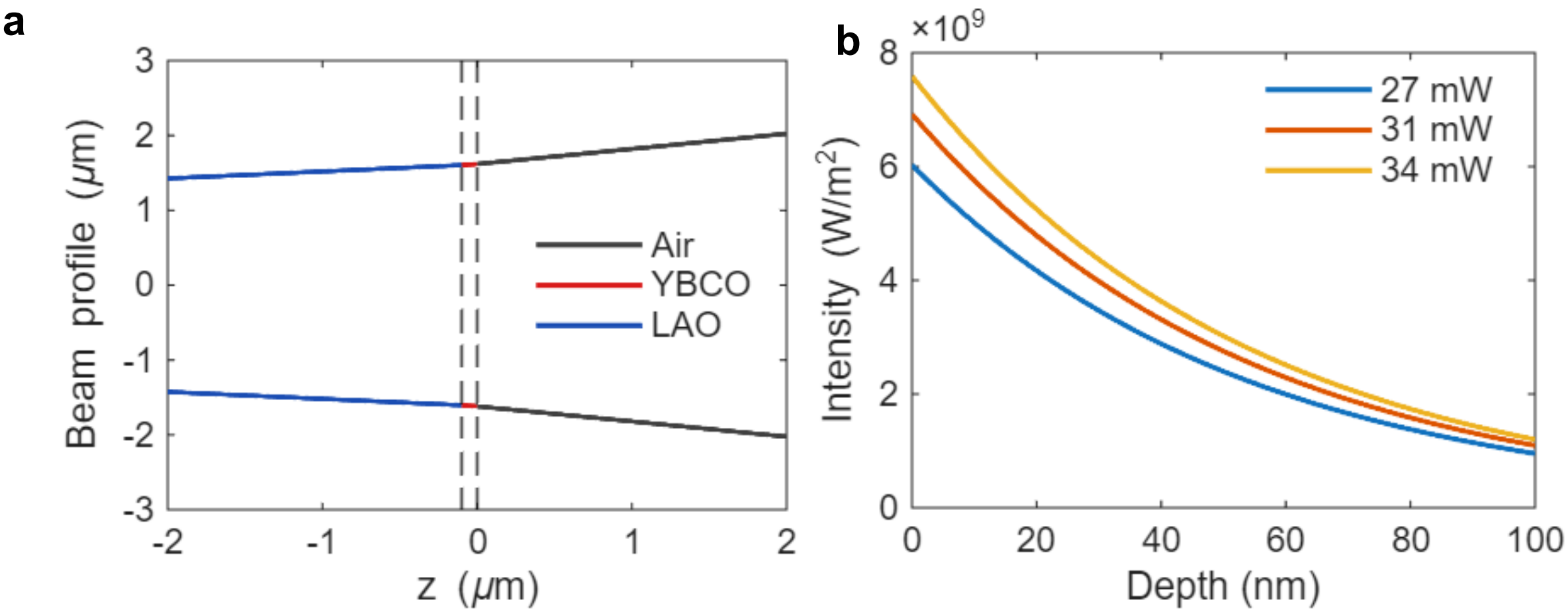

**Figure S1.** (**a**) Calculated laser beam profile across the multilayer structure composed of air, a 100-nm-thick YBCO film, and the LAO substrate, as described in Supporting Note 1. (**b**) Calculated laser intensity profile across the YBCO thickness for incident laser powers of 27, 31, and 34 mW.

**Supporting Note 2**

The oxygen stoichiometry *x* of the $YBa_2Cu_3O_x$ samples was estimated from the Raman shift of the O(4) vibrational mode. This approach is based on well-established results in the literature, which report a linear relationship between the frequency of the O(4) Raman mode and the oxygen content in YBCO over a wide doping range.[3–5]

While this linear dependence allows relative variations in oxygen stoichiometry to be directly inferred from Raman measurements, an absolute determination of *x* requires a reference point with known oxygen content. To this end, we used the pristine sample as a calibration point. The superconducting critical temperature $T_c$ of the pristine sample was independently measured via SQUID magnetometry and serves as a reliable indicator of its oxygen stoichiometry.

The correspondence between $T_c$ and the oxygen content *x* was obtained from a separate literature reference, where experimental data of $T_c$ as a function of *x* for YBCO are reported.[6] Using this established $T_c(x)$ relationship, we extracted the oxygen content of the pristine sample. Once the oxygen stoichiometry of the pristine sample was fixed, the linear relationship between the O(4) Raman shift and *x* was used to determine the oxygen content of all other samples relative to this reference. This procedure allows for a consistent estimation of the oxygen stoichiometry directly from Raman spectroscopy.

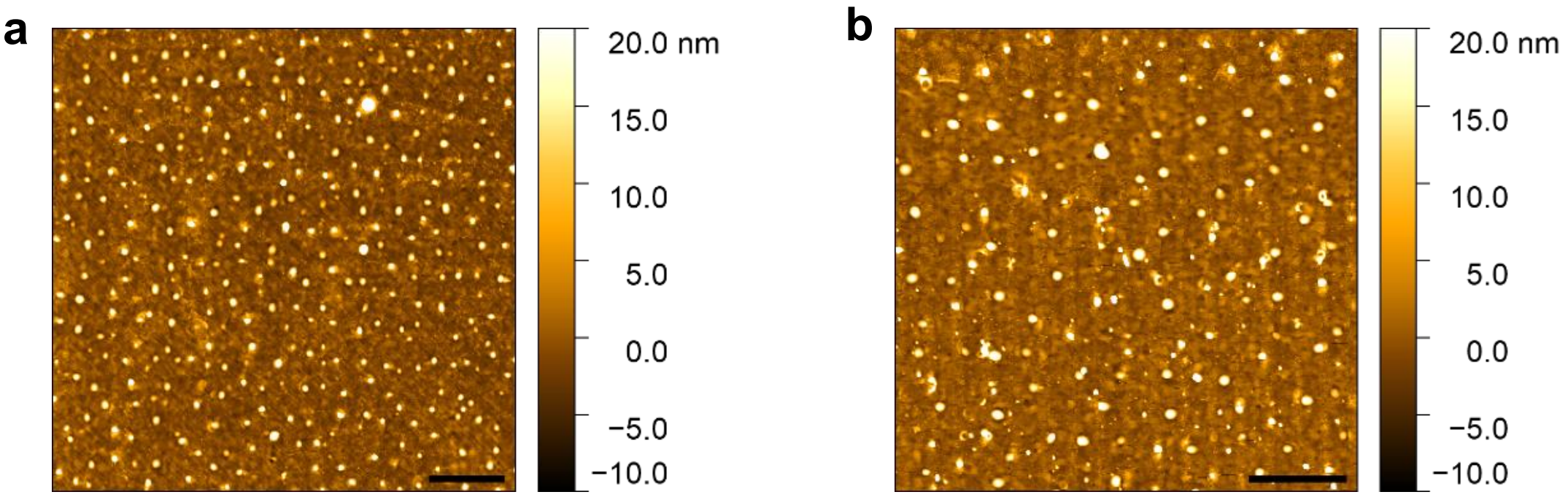

**Figure S2.** Topographic maps measured via atomic force microscopy corresponding to the EFM maps shown in Figure 1d and 1g, respectively. The patterns visible in the EFM images do not produce significant changes in the surface morphology. Scale bars: 2 μm.

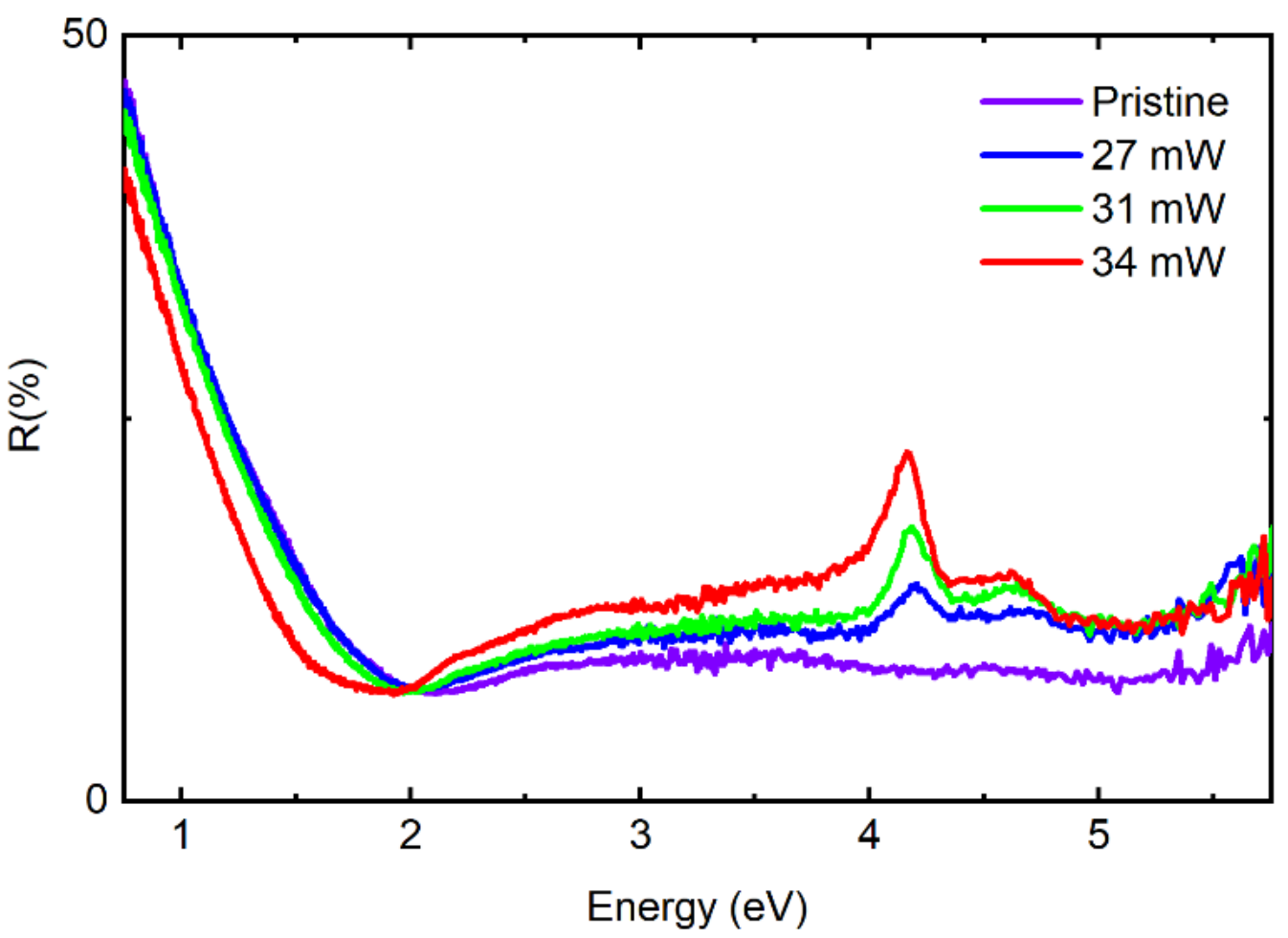

**Figure S3.** Full reflectance spectra of irradiated YBCO squares measured over the 0.75–5.75 eV energy range. This complements Figure 3b, which shows a zoomed-in view of the 3.6–4.8 eV region. While a new peak appears in the zoomed region due to controlled deoxygenation, the overall shape of the spectra remains largely unchanged.

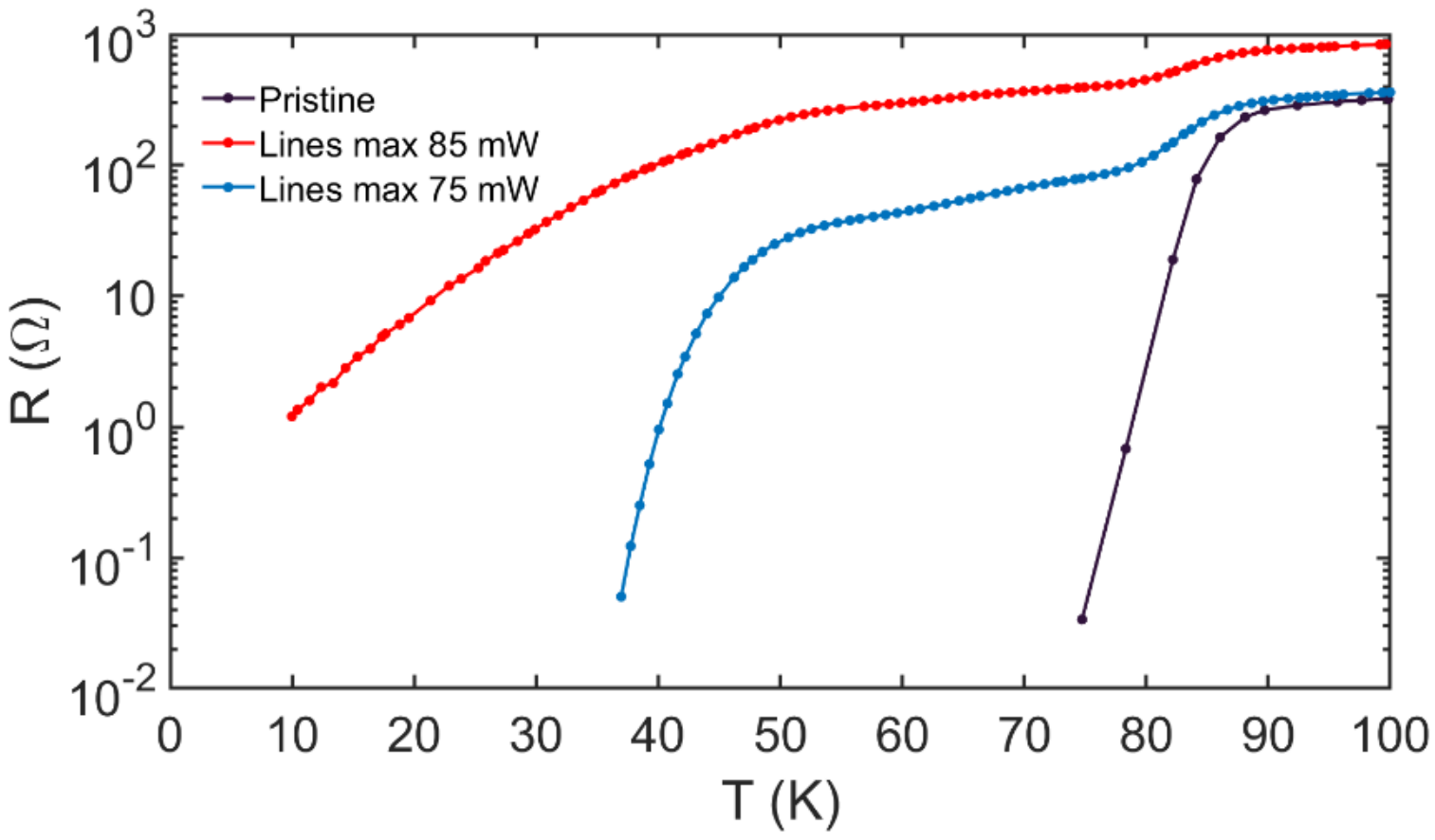

**Figure S4**. Resistance versus temperature measurements of patterned 20-nm-thick YBCO, for a pristine Hall bar and for devices patterned with laser-written lines perpendicular to the current path. In one device, the laser power was varied from 45 to 75 mW in steps of 10 mW, while in a second device powers up to 85 mW were applied. The device containing a line exposed to 85 mW does not exhibit a superconducting transition down to the lowest measured temperature (10 K), unlike the device with a maximum applied power of 75 mW. Instead, it retains a finite residual resistance of approximately 1.2 Ω, indicating a suppression of superconductivity.

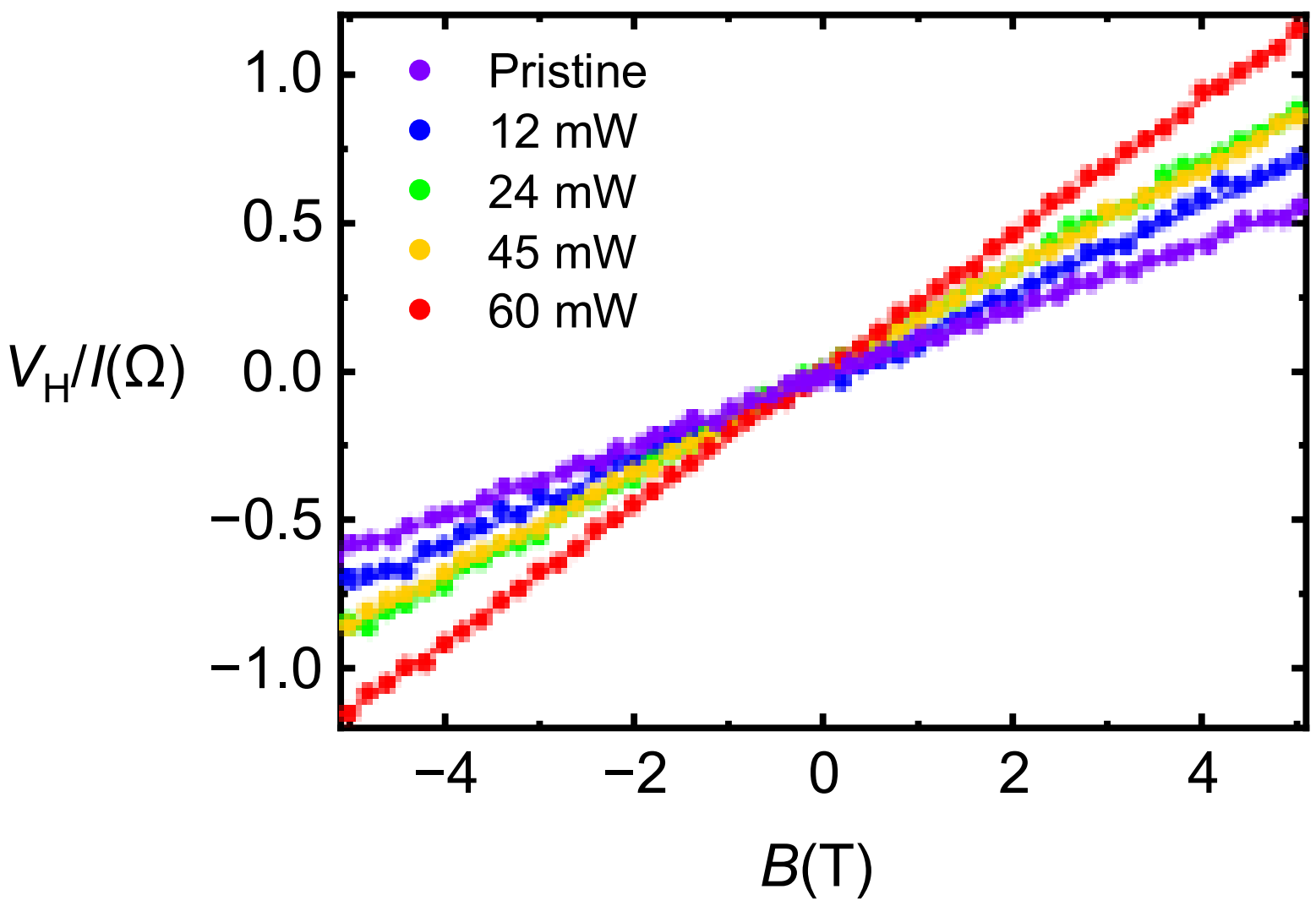

**Figure S5.** Hall voltage normalized by the bias current, $V_{\mathrm{H}}/I$, as a function of the applied perpendicular magnetic field $B$ for the 15-nm-thick YBCO Hall bar devices discussed in Figure 4. Symbols represent the experimental data, while solid lines are linear fits to the high-field regime. The Hall coefficient is determined from the slope of the linear fit according to $R_{\mathrm{H}} = t \times d(V_{\mathrm{H}}/I)/dB$, where $t$ is the film thickness. The Hall carrier density is then calculated as $n_{\mathrm{H}} = 1/(eR_{\mathrm{H}})$, with $e$ being the elementary charge. The obtained $n_{\mathrm{H}}$ values are used to construct the phase diagram shown in Figure 4h.

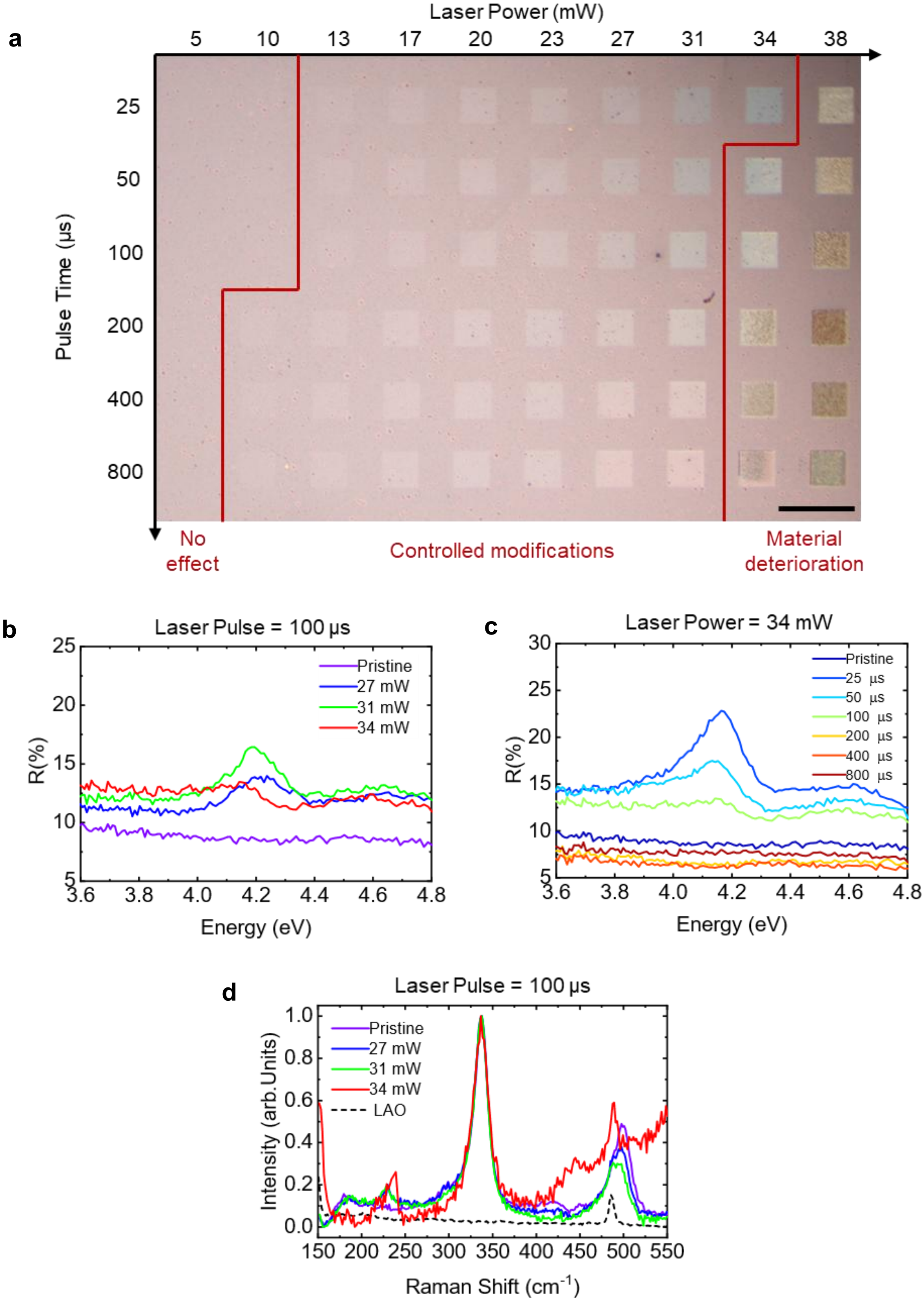

**Figure S6.** Influence of laser exposure time on YBCO patterning. (**a**) Optical image of a matrix of patterned $50 \times 50\ \mu m^2$ squares obtained using different laser powers, from 5 to 38 mW, and pulse durations ranging from 25 to 800 $\mu$s. The laser power required to induce visible modifications, as well as the onset of degradation, depends on the laser pulse duration. Scale bar: 100 $\mu$m. (**b**) Reflectance spectra of patterns written with a pulse duration of 100 $\mu$s at different laser powers. The characteristic feature at 4.1 eV decreases in intensity at the highest applied power (34 mW), unlike in Figure 3b where the spectra acquired for a pulse duration of 25 $\mu$s preserve this feature at all investigated powers. (**c**) Reflectance spectra acquired for patterns written at a fixed power of 34 mW and different pulse durations. The 4.1 eV feature is most pronounced for the shortest exposure time and progressively decreases with increasing pulse duration. (**d**) Raman spectra for patterns written with a pulse duration of 100 $\mu$s and different laser powers. Spectral alterations are observed at 34 mW, whereas lower powers preserve the characteristic Raman features of YBCO. This complements the Raman spectra shown in Figure 3c for a pulse duration of 25 $\mu$s, where the main Raman features are preserved at all investigated powers.

**Movie S1.** Temperature-dependent magneto-optical imaging of laser-patterned YBCO squares. All magneto-optical imaging frames corresponding to Figure 2a-c.

**Movie S2.** Magnetic-field-dependent flux penetration in a laser-patterned YBCO meander. Zero-field-cooled magneto-optical imaging frames corresponding to Figure 2f-j.